\documentclass[pra,reprint,showpacs,groupedaddress,nofootinbib]{revtex4-1}
\usepackage{amsmath,amssymb,amsfonts}
\usepackage{graphicx}
\usepackage{txfonts}

\begin{document}

\title{Variational determination of approximate bright matter-wave soliton solutions in anisotropic traps}

\author{T. P. Billam}
\author{S. A. Wrathmall}
\author{S. A. Gardiner}

\affiliation{Department of Physics, Durham University, Durham DH1 3LE, United Kingdom}

\date{\today}

\pacs{
03.75.Lm      
67.85.Bc      
}

\begin{abstract}
We consider the ground state of an attractively-interacting atomic
Bose-Einstein condensate in a prolate, cylindrically symmetric harmonic trap.
If a true quasi-one-dimensional limit is realized, then for sufficiently weak
axial trapping this ground state takes the form of a bright soliton solution of
the nonlinear Schr\"{o}dinger equation.  Using analytic variational and highly
accurate numerical solutions of the Gross-Pitaevskii equation we systematically
and quantitatively assess how soliton-like this ground state is, over a wide
range of trap and interaction strengths. Our analysis reveals that the regime
in which the ground state is highly soliton-like is significantly restricted,
and occurs only for experimentally challenging trap anisotropies. This result,
and our broader identification of regimes in which the ground state is
well-approximated by our simple analytic variational solution, are relevant to
a range of potential experiments involving attractively-interacting
Bose-Einstein condensates.
\end{abstract}

\maketitle

\section{Introduction}
Bright solitons are self-focusing, non-dispersive, particle-like solitary waves
occurring in integrable systems \cite{dauxois_book, fadeev}. They behave in a
particle-like manner, emerging from mutual collisions intact except for shifts
in their position and relative phase.  Bright soliton solutions of the
one-dimensional nonlinear Schr\"{o}dinger equation (NLSE) can be described
analytically using the inverse scattering technique \cite{zakharov_shabat_1972,
satsuma_yajima_1974} and are well-known in the context of focusing
nonlinearities in optical fibers \cite{satsuma_yajima_1974, gordon_ol_1983}.
Bright solitary matter-waves in an attractively interacting atomic
Bose-Einstein condensate (BEC) represent an intriguing alternative physical
realization \cite{khaykovich_etal_science_2002, strecker_etal_nature_2002,
cornish_etal_prl_2006}. In a mean-field description an atomic BEC obeys the
Gross-Pitaevskii equation (GPE) \cite{pitaevskii_stringari_2003}, a
three-dimensional NLSE. While in general non-integrable, in a homogeneous,
quasi-one-dimensional (quasi-1D) limit the GPE reduces to the one-dimensional
NLSE, thus supporting bright solitons \cite{perez-garcia_etal_prl_1996,
perez-garcia_etal_pra_1998, salasnich_etal_pra_2002b, salasnich_etal_pra_2002a,
martin_etal_prl_2007, martin_etal_pra_2008}. 

Outside the quasi-1D limit the GPE continues to support bright solitary
matter-waves. These exhibit many soliton-like characteristics and have been the
subject of much experimental \cite{khaykovich_etal_science_2002,
strecker_etal_nature_2002, cornish_etal_prl_2006} and theoretical
\cite{parker_etal_jpb_2008, parker_etal_physicad_2008, billam_etal_pra_2011,
poletti_etal_physicad_2009, cornish_etal_physicad_2009,
khaykovich_malomed_pra_2006, cornish_etal_physicad_2009,
al_khawaja_etal_prl_2002, strecker_etal_njp_2003, carr_brand_prl_2004,
dabrowska_wuster_etal_njp_2009, al_khawaja_stoof_njp_2011,
weiss_castin_prl_2009, streltsov_etal_prl_2007, streltsov_etal_pra_2009}
investigation.  Both bright solitons and bright solitary waves are excellent
candidates for use in atom interferometry \cite{cronin_etal_rmp_2009}, as their
coherence, spatial localization and soliton-like dynamics offer a metrological
advantage in, e.g., the study of atom-surface interactions
\cite{strecker_etal_nature_2002, cornish_etal_physicad_2009}. Towards this end,
proposals to phase-coherently split bright solitons and bright solitary waves
using a scattering potential \cite{streltsov_etal_prl_2007,
streltsov_etal_pra_2009, weiss_castin_prl_2009} and an internal state
interference protocol \cite{billam_etal_pra_2011}, and to form soliton
molecules \cite{al_khawaja_stoof_njp_2011} have been explored in the
literature. However, while the dynamics and collisions of bright solitary waves
have been explored in detail and have been shown to be soliton-like in
three-dimensional (3D) parameter regimes \cite{parker_etal_jpb_2008,
parker_etal_physicad_2008, billam_etal_pra_2011}, less attention has been
directed at the question of exactly how soliton-like the ground state of the
system is. In particular, the experimental feasibility of reaching the quasi-1D
limit of an attractively-interacting BEC, and hence obtaining a highly
soliton-like ground state, remains an area lacking a thorough quantitative
exploration. Obtaining such a ground state, in addition to being interesting in
its own right, would be highly advantageous in experiments seeking to probe
quantum effects beyond the mean-field description
\cite{streltsov_etal_prl_2007, streltsov_etal_pra_2009, weiss_castin_prl_2009},
and possibly to exploit the effects of macroscopic quantum superposition to
enhance metrological precision \cite{dunningham_burnett_pra_2004,
dunningham_contemp_phys_2006}. Similar concerns regarding adverse residual 3D
effects in interferometric protocols prompted a recent perturbative study of
residual 3D effects in highly anisotropic, repulsively-interacting BECs
\cite{tacla_calves_pra_2011}.

The potential instability to collapse of attractively-interacting BECs
\cite{parker_etal_jpb_2007, ruprecht_etal_pra_1995, roberts_etal_prl_2001,
donley_etal_nature_2001, gerton_etal_nature_2000, dodd_etal_pra_1996, pitaevskii_pla_1996,
savage_etal_pra_2003, altin_etal_pra_2011, gammal_etal_pra_2001,
gammal_etal_pra_2002} is the key obstacle to realizing soliton-like behavior in
a BEC.  Previous studies of bright solitary wave dynamics, using variational
and numerical solutions of partially-quasi-1D GPEs
\cite{khaykovich_malomed_pra_2006, salasnich_etal_pra_2002b,
salasnich_etal_pra_2002a, kamchatnov_shchesnovich_pra_2004} [reductions of the
GPE to a 1D equation which retain some 3D character, in contrast to the full
quasi-1D limit] and the 3D GPE \cite{parker_etal_jpb_2007,
parker_etal_jpb_2008, parker_etal_physicad_2008, billam_etal_pra_2011}, have
shown the collapse instability to be associated with non-soliton-like behavior.
However, previous studies of bright solitary wave ground states have focused on
identifying the critical parameters at which collapse occurs.  Approaches used
in these studies include partially-quasi-1D methods
\cite{salasnich_etal_pra_2002b}, variational methods \cite{malomed_po_2002} using Gaussian
\cite{perez-garcia_etal_prl_1996, perez-garcia_etal_pra_1997,
parker_etal_jpb_2007} and soliton (sech)
\cite{parker_etal_jpb_2007,carr_castin_pra_2002} ansatzes, perturbative methods
\cite{yukalov_yukalova_pra_2005}, and numerical solutions to the 3D GPE
\cite{ruprecht_etal_pra_1995, parker_etal_jpb_2007, gammal_etal_pra_2001,
gammal_etal_pra_2002, carr_castin_pra_2002}. In the latter case, the collapse
threshold parameters have been extensively mapped out for a range of trap
geometries \cite{gammal_etal_pra_2001, gammal_etal_pra_2002}.

In this paper we use analytic variational and highly accurate numerical
solutions of the stationary GPE to systematically and quantitatively assess how
soliton-like the ground state of an attractively-interacting BEC in a prolate,
cylindrically symmetric harmonic trap is, over a wide regime of trap and
interaction strengths.  Beginning with previously-considered variational
ansatzes based on Gaussian \cite{perez-garcia_etal_prl_1996,
perez-garcia_etal_pra_1997, parker_etal_jpb_2007} and soliton
\cite{parker_etal_jpb_2007,carr_castin_pra_2002} profiles, we obtain new,
analytic variational solutions for the GPE ground state. Comparing the
soliton-ansatz variational solution to highly accurate numerical solutions of
the stationary GPE, which we calculate over an extensive parameter space, gives
a quantitative measure of how soliton-like the ground state is. In the regime
where the axial and radial trap strengths dominate over the interactions, we
show that the Gaussian ansatz variational solution gives an excellent
approximation to the true ground state for all anisotropies; in this regime the
ground state is not soliton-like.  In the regime in which the interactions
dominate over the axial, but \textit{not} the radial, trap strength we
demonstrate that the soliton-ansatz variational solution does approximate the
true, highly soliton-like ground state.  However, we show that the goodness of
the approximation and the extent of this regime, where it exists at all, is
highly restricted by the collapse instability; even at large anisotropies it
occupies a narrow window adjacent to the regime where interactions begin to
dominate over \textit{all} trap strengths, leading to non-quasi-1D,
non-soliton-like solutions and, ultimately, collapse.  

Our results have substantial practical value for experiments using
attractively-interacting BECs; primarily they define the challenging
experimental regime required to realize a highly soliton-like ground state,
which would be extremely useful to observe quantum effects beyond the
mean-field description such as macroscopic superposition of solitons
\cite{streltsov_etal_prl_2007, streltsov_etal_pra_2009, weiss_castin_prl_2009}.
We note that bright solitary wave experiments to date have not reached this
regime \cite{khaykovich_etal_science_2002, strecker_etal_nature_2002,
cornish_etal_prl_2006}. Secondarily, our quantitative analysis of a wide
parameter space provides a picture of the ground state in a wide range of
possible attractively-interacting BEC experiments. In particular, it indicates
the regimes in which a full numerical solution of the 3D GPE is
well-approximated by one of our analytic variational solutions, which are
significantly easier and less time-consuming to determine.

The remainder of the paper is structured as follows: After introducing the most
general classical field Hamiltonian and stationary GPE in Section
\ref{s_3dgpe}, we begin by discussing the quasi-1D limit in Section
\ref{s_quasi1dlimit}. In Section \ref{s_1dlimit} we define the dimensionless
trap frequency $\gamma$; in the quasi-1D limit this is the \textit{only} free
parameter, and all our results are expressed in terms of this quantity.
Similarly, our variational ansatzes are motivated by the limiting behaviors of
the solution in the quasi-1D case; in this case we define them as Gaussian and
soliton profiles, parametrized by their axial lengths. In Sections
\ref{s_1dgauss} and \ref{s_1dsol} we find, analytically, the energy-minimizing
axial lengths for each ansatz as a function of $\gamma$. Comparison of the
resulting ansatz solutions to highly accurate numerical solutions of the
stationary quasi-1D GPE allows us to determine, in the quasi-1D limit, the
regimes of low $\gamma$ in which highly soliton-like ground states can be
realized (Section \ref{s_1dcompare}). We then consider the 3D GPE in Section
\ref{s_3d}.  The system then has a second free parameter in addition to
$\gamma$; we choose this to be $\kappa$, the (dimensionless) trap anisotropy,
which is defined in Section \ref{s_3dlimit}. In Sections \ref{s_3dstart} to
\ref{s_3dend} we define 3D Gaussian and soliton ansatzes, adapted from their
quasi-1D analogs and each parametrized by an axial and a radial length, and
find the energy-minimizing lengths for each ansatz. In general this requires
only a very simple numerical procedure, and in the limit of a waveguide-like
trap can be expressed analytically (Section \ref{s_waveguide}). In Section
\ref{s_3dcompare} we compare the ansatz solutions to highly accurate numerical
solutions of the stationary 3D GPE and, in Section \ref{s_discuss} assess the
potential for realizing truly soliton-like ground states. Finally, Section
\ref{s_conclusions} comprises the conclusions.

\section{System overview\label{s_3dgpe}}
We consider a BEC of $N$ atoms of mass $m$ and (attractive) $s$-wave scattering
length $a_{s}<0$, held within a cylindrically symmetric, prolate (the radial
frequency $\omega_{r}$ is greater than the axial frequency $\omega_x$) harmonic
trap. The ground state is described by the stationary Gross-Pitaevskii equation
\begin{equation}
 \left[
 - \frac{\hbar^2}{2m} \nabla^{2} + V(\mathbf{r}) - \frac{4 \pi N |a_s| \hbar^2}{m} |\psi(\mathbf{r})|^{2} -\lambda \right]
 \psi(\mathbf{r}) = 0,
\label{Eq:3DGPE}
\end{equation} 
where the trapping potential $V(\mathbf{r})=m[\omega_{x}^{2}
x^{2}/2+\omega_{r}^{2} (y^{2}+z^{2})/2]$, $\lambda$ is a real eigenvalue, and
the Gross-Pitaevskii wavefunction $\psi(\mathbf{r})$ is normalized to one.
This equation is generated by the classical field Hamiltonian (through the
functional derivative $ \delta H[\psi]/\delta \psi^{*} = \lambda \psi$)
\begin{multline}
H[\psi] = \int d\mathbf{r} \left[
\frac{\hbar^{2}}{2m}\left|\nabla\psi(\mathbf{r})\right|^{2}
+V(\mathbf{r})|\psi(\mathbf{r})|^{2} 
\right. \\ \left.
-\frac{2\pi N|a_s|\hbar^2}{m}|\psi(\mathbf{r})|^{4}
\right].
\label{Eq:GeneralFunctional}
\end{multline}
This functional of the classical field $\psi$ describes the total energy per
particle, and the ground state solution minimizes the value of this functional. 

When dealing with variational ansatzes for the ground state solution, we
proceed by analytically minimizing an energy functional in the same form as
Eq.\ (\ref{Eq:GeneralFunctional}) for a given ansatz. In contrast, highly
accurate numerical ground states are more conveniently obtained by solving a
stationary GPE of the same form as Eq.\ (\ref{Eq:3DGPE}).

\section{Quasi-1D limit\label{s_quasi1dlimit}}
\subsection{Reduction to 1D and rescaling\label{s_1dlimit}}
For sufficiently tight radial confinement ($\omega_{r}\gg\omega_{x}$), such
that the atom-atom interactions are nonetheless essentially 3D [$a_{s}\ll
(\hbar/m\omega_{r})^{1/2}$] it is conventional
\cite{perez-garcia_etal_prl_1996, perez-garcia_etal_pra_1998,
salasnich_etal_pra_2002b, salasnich_etal_pra_2002a, martin_etal_prl_2007,
martin_etal_pra_2008} to assume a reduction to a quasi-1D stationary GPE
\begin{equation}
 \left[
 - \frac{\hbar^2}{2m} \frac{\partial^{2}}{\partial x^{2}} + \frac{m\omega_x^{2}x^{2}}{2} - g_{\textrm{1D}}N|\psi(x)|^{2} - \lambda \right]
 \psi(x) = 0.
\label{Eq:1DGPE}
\end{equation} 
Typically $\psi(\mathbf{r})$ is taken to be factorized into $\psi(x)$ and the
radial harmonic ground state $(m\omega_{r}/\pi\hbar)^{1/2}
\exp(-m\omega_{r}[y^{2}+z^{2}]/2\hbar)$, such that $g_{\textrm{1D}} =
2\hbar\omega_{r}|a_{s}|$. Alternative factorizations are also possible, which
lead to an effective 1D equation retaining more 3D character than Eq.\
(\ref{Eq:1DGPE}) \cite{khaykovich_malomed_pra_2006, salasnich_etal_pra_2002b,
salasnich_etal_pra_2002a, kamchatnov_shchesnovich_pra_2004}; similar
factorizations have also been introduced for axially rotating BECs \cite{salasnich_etal_pra_2007} and
for quasi-2D BECs in oblate traps \cite{salasnich_etal_pra_2009}. In the absence of the axial harmonic
confining potential ($\omega_x \rightarrow 0$), there exist exact bright
soliton solutions to Eq.\ (\ref{Eq:1DGPE}) of the general
form\footnote{Equation (\ref{Eq:Solitons}) describes solutions of unit norm.
More general soliton solutions $(B/2b_{x}^{1/2}) \mbox{sech}(B[x - vt +
C]/2b_{x}) e^{iv(x-vt)m/\hbar} e^{iB^{2}mg_{\textrm{1D}}^{2}N^2t/8\hbar^{3}}
e^{imv^{2}t/2\hbar} e^{iD}$ have norm $B$ (and effective mass $\eta=B/4$), as
arise when considering several solitons simultaneously.} 
\begin{equation}
\frac{1}{2 b_{x}^{1/2}} \mbox{sech}\left(\frac{[x - vt + C]}{2b_{x}}\right)
e^{iv(x-vt)m/\hbar} e^{img_{\textrm{1D}}^{2}N^2t/8\hbar^{3}}
e^{imv^{2}t/2\hbar} e^{iD}, 
\label{Eq:Solitons} 
\end{equation} 
where $b_{x}\equiv\hbar^{2}/mg_{\textrm{1D}}N$ is a length scale characterizing
the soliton's spatial extent, $v$ is the soliton velocity, $C$ is an arbitrary
displacement, and $D$ is an arbitrary phase.

This effective 1D Gross-Pitaevskii equation contains two key length scales: the
axial harmonic length $a_{x} \equiv (\hbar/m\omega_{x})^{1/2}$, and the soliton
length $b_{x}$. A mathematically convenient way to express the single free
parameter of Eq.\ (\ref{Eq:1DGPE}) is as the square of the ratio of these two
length scales;
\begin{equation}
\gamma \equiv \left(\frac{b_{x}}{a_{x}}\right)^{2} \equiv \frac{\hbar \omega_x}{4m\omega_r^{2}|a_s|^2N^2}.
\end{equation}
This parametrization is achieved by working in ``soliton units'';
lengths are expressed in units of $b_{x}$ and energies are expressed in units of
$mg_{\textrm{1D}}^{2}N^2/\hbar^2$. This system can be codified as $\hbar = m =
g_{\textrm{1D}}N = 1$, and yields the dimensionless quasi-1D GPE
\begin{equation}
 \left[
 - \frac{1}{2} \frac{\partial^{2}}{\partial x^{2}} + \frac{\gamma^{2}x^{2}}{2} -|\psi(x)|^{2} -\lambda \right]
 \psi(x) = 0,
\label{Eq:Scaled1DGPE}
\end{equation}
in which $\gamma$ can be interpreted as a dimensionless trap frequency \cite{martin_etal_pra_2008}. 
The corresponding classical field Hamiltonian is
\begin{equation}
H_{\textrm{1D}}[\psi] = \int dx \left[
\frac{1}{2}\left|\frac{\partial }{\partial x}\psi(x)\right|^{2}
+\frac{\gamma^{2}x^{2}}{2}|\psi(x)|^{2} 
-\frac{1}{2}|\psi(x)|^{4} \right].
\label{Eq:1DFunctional}
\end{equation}

The choice of $\gamma$ for the single free parameter in the 1D GPE [Eq.\
(\ref{Eq:Scaled1DGPE})] and the classical field Hamiltonian [Eq.\
(\ref{Eq:1DFunctional})] can be most directly pictured as choosing to hold
interaction strength constant while varying the axial trap strength,
parametrized by $\gamma$. Experimentally, however, any of $\omega_x$,
$\omega_r$, $a_s$, and $N$ may be varied in order to vary $\gamma$. In the case
$\gamma=0$ the exact ground state solution is a single, stationary bright
soliton: $\psi(x) = \mbox{sech}(x/2)/2$. In the following subsections we
develop analytic variational solutions $\psi(x)$ for general $\gamma$.
Comparing these solutions to highly accurate numerical solutions of the
quasi-1D GPE then gives a picture of the behavior of the ground state with
$\gamma$. Furthermore, these quasi-1D variational solutions motivate the later
3D variational solutions and yield several mathematical expressions
which reappear in the more complex 3D calculations.

\subsection{Variational solution: Gaussian ansatz\label{s_1dgauss}}
We first consider the Gaussian variational ansatz 
\begin{equation}
\psi(x) = \left(\frac{\gamma}{\pi \ell_{\textrm{G}}^{2}}\right)^{1/4} e^{-\gamma x^{2}/2 \ell_\textrm{G}^2},
\label{Eq:1DGauss}
\end{equation}
where the variational parameter, $\ell_\textrm{G}$, quantifies the axial
length.  In the trap-dominated limit ($\gamma \rightarrow \infty$), the true
solution tends to a Gaussian with $\ell_\textrm{G} = 1$.  Substituting Eq.\
(\ref{Eq:1DGauss}) into Eq.\ (\ref{Eq:1DFunctional}) yields (using identities
from Appendix \ref{appendix_integrals})
\begin{equation}
H_{\textrm{1D}}(\ell_{\textrm{G}}) = 
\frac{\gamma}{4} \left(\ell_\textrm{G}^2 + \frac{1}{\ell_\textrm{G}^2} - \frac{2}{(2\pi\gamma)^{1/2}\ell_\textrm{G}} \right),
\label{Eq:GaussVariational}
\end{equation}
where $H_{\textrm{1D}}$ is now expressed as a function of the axial length
 $\ell_{\textrm{G}}$. Setting $\partial H_{\textrm{1D}}/\partial
\ell_{\textrm{G}}=0$ reveals that the variational energy described by Eq.\
(\ref{Eq:GaussVariational}) is minimized when $\ell_{\textrm{G}}$ is a
positive, real solution to the quartic equation
\begin{equation}
\ell_{\textrm{G}}^{4} + \frac{\ell_{\textrm{G}}}{(2\pi\gamma)^{1/2}} -1 = 0.
\end{equation}
The positive, real solution to this quartic is (see solution in Appendix
\ref{appendix_quartic})
\begin{equation}
\ell_{\textrm{G}} = 
\frac{\left[\chi(\gamma)\right]^{1/2}}{2^{4/3}(\pi\gamma)^{1/6}}\left\{
\left[
\left(\frac{2}{\chi(\gamma)}\right)^{3/2} - 1
\right]^{1/2}  -1\right\},
\label{equation_1d_gauss_sol}
\end{equation}
where we have, for notational convenience, defined $\chi$ to have
$\gamma$-dependence such that
\begin{multline}
\chi(\gamma) =
\left[
1+
\left(
1 + \frac{1024\pi^{2}\gamma^{2}}{27}
\right)^{1/2}
\right]^{1/3}
\\ 
+\left[
1-
\left(
1 + \frac{1024\pi^2\gamma^{2}}{27}
\right)^{1/2}
\right]^{1/3}.
\label{equation_chi_def}
\end{multline}

\subsection{Variational solution: soliton ansatz\label{s_1dsol}}
Secondly, we consider a soliton ansatz
\begin{equation}
\psi(x) = \frac{1}{2\ell_{\textrm{S}}^{1/2}}\mbox{sech}\left(\frac{x}{2\ell_{\textrm{S}}}\right),
\end{equation}
where the variational parameter, $\ell_\textrm{S}$, again quantifies the axial
length.  In the axially un-trapped limit ($\gamma \rightarrow 0$),  the true
solution tends to a classical bright soliton, as described by the above ansatz
with  $\ell_{\textrm{S}} = 1$. The variational energy per particle is given by
(using identities from Appendix \ref{appendix_integrals})
\begin{equation}
H_{\textrm{1D}}(\ell_{\textrm{S}}) = 
\frac{\pi^2\gamma^2}{6}\left(\ell_{\textrm{S}}^{2} + \frac{1}{4 \pi^2 \gamma^2 \ell_{\textrm{S}}^{2}} -\frac{1}{2 \pi^2 \gamma^2 \ell_{\textrm{S}}} \right),
\end{equation}
which is minimized when
\begin{equation}
\ell_{\textrm{S}}^{4}
+ \frac{\ell_{\textrm{S}}}{4\pi^2\gamma^{2}}
- \frac{1}{4\pi^2\gamma^{2}} = 0.
\end{equation}
Again, this quartic can be solved analytically (see solution in
Appendix \ref{appendix_quartic}) to give the positive, real minimizing value
of $\ell_{\textrm{S}}$;
\begin{equation}
\ell_{\textrm{S}} = \frac{\left[\chi(\gamma)\right]^{1/2}}{2^{11/6}(\pi\gamma)^{2/3}}
\left\{
\left[
\left(
\frac{2}{\chi(\gamma)}
\right)^{3/2}
-1
\right]^{1/2}
-1
\right\},
\label{equation_1d_sech_sol}
\end{equation}
with $\chi$ defined as in Eq.\ (\ref{equation_chi_def}).

\subsection{Analysis and comparison to 1D numerical solutions\label{s_1dcompare}}
\begin{figure}
\includegraphics[width=3.375in]{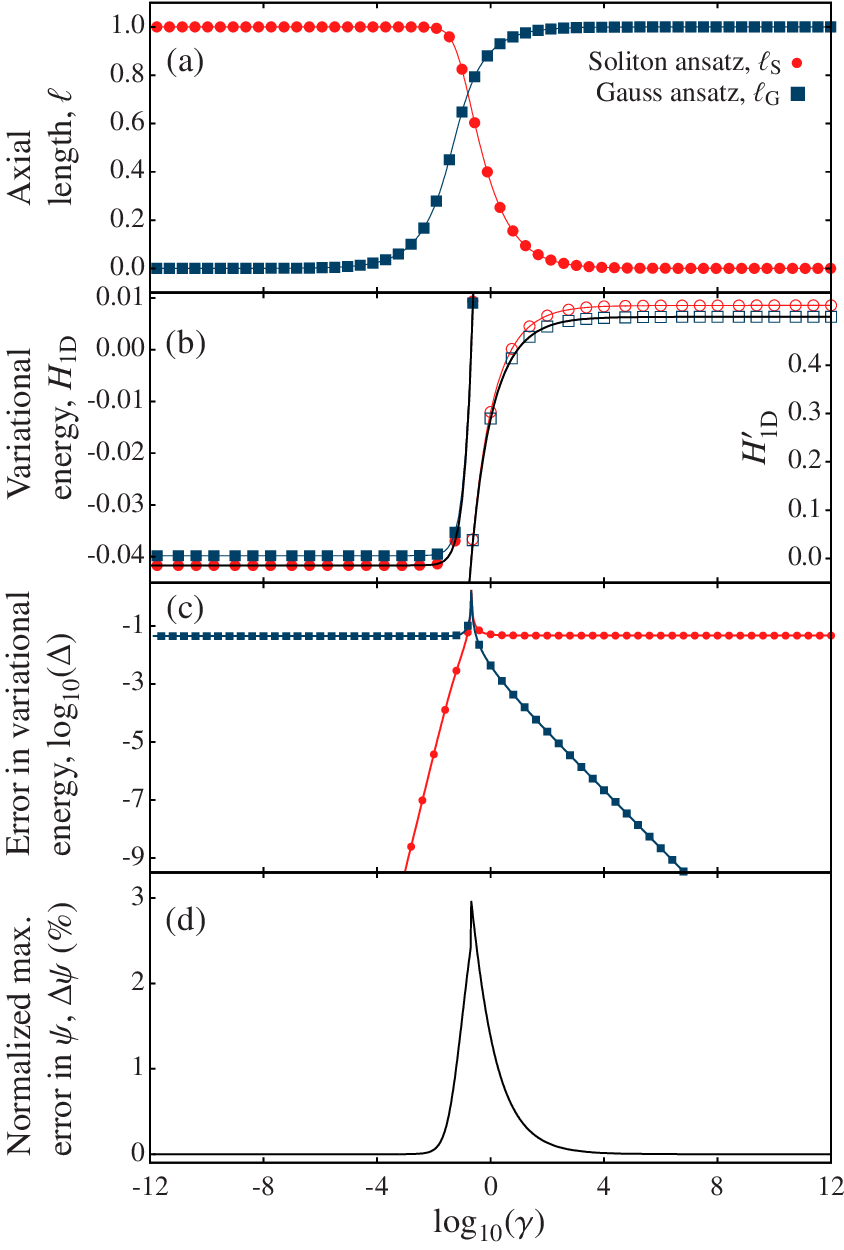}
\caption{Comparison of quasi-1D variational and numerical solutions: (a)
Energy-minimizing axial lengths $\ell_\textrm{G}$ (Gaussian ansatz, squares)
and $\ell_\textrm{S}$ (soliton ansatz, circles) for the quasi-1D GPE. (b)
Minimum variational energy compared with the numerically calculated ground
state energy $E_\textrm{1D}$ (black line) for each ansatz: for low $\gamma$ we
show $H_\textrm{1D}$ (solid symbols), which tends to $-1/24$ as
$\gamma\rightarrow 0$; for high $\gamma$ we show $H_\textrm{1D}^\prime
=H_\textrm{1D}/\gamma$ (hollow symbols), which tends to $1/2$ as $\gamma
\rightarrow \infty$ ($H_\textrm{1D}^\prime$ is equal to the energy expressed in
the ``harmonic units,'' $\hbar = m = \omega_x = 1$).  (c) Relative error in the
variational energy, $\Delta = (H_\textrm{1D} - E_\textrm{1D})/E_\textrm{1D}$.
(d) Normalized maximum deformation of the best-fitting ansatz wavefunction
$\psi_\mathrm{Ansatz}$ with respect to the numerical ground state $\psi_0$,
$\Delta\psi = \mathrm{max}(|\psi_\mathrm{Ansatz} - \psi_0|) /
\mathrm{max}(\psi_0)$, expressed as a percentage.  For clarity in (a,b) [(c)],
every 16th [20th] datum is marked by a symbol.}
\label{figure_energies_1d}
\end{figure}

The energy-minimizing axial lengths $\ell_\textrm{G}$ and $\ell_\textrm{S}$,
defined by Eq.\ (\ref{equation_1d_gauss_sol}) and Eq.\
(\ref{equation_1d_sech_sol}) respectively, are shown as a function of $\gamma$
in Fig.\ \ref{figure_energies_1d}(a).  There is no collapse instability in the
quasi-1D GPE, and solutions are obtained for all (positive, real) $\gamma$. As
intended by the chosen forms of the ansatzes, the limiting cases are
$\ell_\textrm{G} \rightarrow 1$ as $\gamma \rightarrow \infty$ and
$\ell_\textrm{S} \rightarrow 1$ as $\gamma \rightarrow 0$.  To evaluate the
accuracy of the ansatzes for general $\gamma$, we compare each ansatz with the
numerically determined ground state of the quasi-1D GPE. The computation of a
numerically exact ground state $\psi_0(x)$, and the corresponding ground state
energy $E_\textrm{1D}$, uses a pseudospectral method in a basis of symmetric
Gauss-Hermite functions; this is a simplified version of the pseudospectral
method used for 3D calculations, which is explained in more detail in the next
section. Several quantities are compared in Fig.\
\ref{figure_energies_1d}(b--d): the variational minimum energies
$H_\textrm{1D}$ for each ansatz and the numerical ground state energy
$E_\textrm{1D}$ are shown in Fig.\ \ref{figure_energies_1d}(b); the relative
error between $H_\textrm{1D}$ and $E_\textrm{1D}$, defined as $\Delta =
(H_\textrm{1D}-E_\textrm{1D})/|E_\textrm{1D}|$, is shown for each ansatz in
Fig.\ \ref{figure_energies_1d}(c); and the maximum difference between the most
appropriate ansatz wavefunction (that with lowest $\Delta$) and the numerical
ground state wavefunction, expressed as a percentage of the maximum value of
the numerically exact ground state, $\Delta \psi =
\textrm{max}(|\psi_\textrm{Ansatz}-\psi_0|)/\textrm{max}(\psi_0)$ [Fig.\
\ref{figure_energies_1d}(d)].  All the shown computed quantities are
insensitive to a doubling of the numerical basis size from 500 to 1000 states.

Both the Gaussian and soliton ansatzes provide an excellent approximation to
the exact solutions over a large range of $\gamma$. In the regimes where the
relative error in the energy $\Delta$ becomes significantly lower than
$10^{-9}$ in particular, the difference between the ansatz solutions and
numerical solutions becomes generally indistinguishable from numerical
round-off error. For the Gaussian ansatz the convergence to this regime is
noticeably slower than for the soliton ansatz [Fig.\
\ref{figure_energies_1d}(c)]. This effect is a consequence of the
parametrization in terms of $\gamma$ and the corresponding ``soliton units'':
increasing $\gamma$ leads not only to to higher trap strength, but also to
higher peak densities $|\psi(x)|^2$, and hence a stronger nonlinear effect.

For later comparison to the 3D case, it is useful to define a benchmark value
of the relative error $\Delta$ that indicates excellent agreement between the
ansatz and the numerically exact solution. Such a definition, however, will
vary according to purpose. As our objectives in this paper relate significantly
to the \textit{shape} of the ground state, this forms the basis of our
benchmark; a maximum deformation of the wavefunction below 0.1\% of the peak
value [as measured by $\Delta \psi$ in Fig.\ \ref{figure_energies_1d}(d)]
corresponds very closely to $\Delta<10^{-5}$.  Because the relative error
$\Delta$ saturates to a background value of $\approx 10^{-1}$ in regimes where
the chosen ansatz is inapplicable, a value of $\Delta$ four orders of magnitude
below this background value thus corresponds to an excellent match in shape
between the ansatz and the numerically exact solution. With respect to this
benchmark, the Gaussian ansatz represents an excellent fit for
$\log_{10}(\gamma) > 1.15$, while the ground state is highly soliton-like (the
soliton ansatz represents an excellent fit) for $\log_{10}(\gamma) < -0.95$.

\section{Bright solitary wave ground states in 3D\label{s_3d}}
\subsection{Rescaling to effective 1D soliton units\label{s_3dlimit}}
We now consider the cylindrically symmetric 3D Gross-Pitaevskii equation [Eq.\
(\ref{Eq:3DGPE})].  Compared to the quasi-1D effective Gross-Pitaevskii
equation [Eq.\ (\ref{Eq:Scaled1DGPE})], three-dimensionality introduces an
additional relevant length scale, the radial harmonic length
$a_{r}=(\hbar/m\omega_{r})^{1/2}$. We incorporate this into the dimensionless
trap anisotropy $\kappa \equiv \omega_{r}/\omega_{x}$, which forms an
additional free parameter.  Expressed in the same ``soliton units'' as Eq.\
(\ref{Eq:Scaled1DGPE}), Eq.\ (\ref{Eq:3DGPE}) becomes
\begin{equation}
 \left[
 - \frac{1}{2} \nabla^{2} 
 + V(\mathbf{r}) 
 - \frac{2\pi}{\kappa\gamma} |\psi(\mathbf{r})|^{2} - \lambda \right]
 \psi(\mathbf{r}) = 0,
\label{Eq:Scaled3DGPE}
\end{equation}
with corresponding energy functional
\begin{multline}
H_{\textrm{3D}}[\psi] = \int d\mathbf{r} \left[
\frac{1}{2}\nabla\psi(\mathbf{r})\cdot \nabla\psi^{*}(\mathbf{r}) \right.\\
+\left. V(\mathbf{r})|\psi(\mathbf{r})|^{2} - \frac{\pi}{\kappa\gamma}|\psi(\mathbf{r})|^{4}
\right],
\label{Eq:3DFunctional}
\end{multline}
 where $V(\mathbf{r}) = \gamma^{2}[x^{2} + \kappa^{2}(y^{2}+z^{2})]/2$.

In the following subsections we obtain variational solutions for general
$\kappa$ and $\gamma$ using ansatzes similar to the Gaussian and soliton
ansatzes employed in the previous section, with an additional variable-width
Gaussian radial profile. Contrary to the case in the quasi-1D limit, a
self-consistent energy-minimizing solution for both the axial and radial length
parameters cannot be expressed entirely analytically. However, we reduce the
numerical work required to the simultaneous solution of two equations, and
introduce a straightforward iterative technique to achieve this. We also
consider the case of a waveguide-like trap ($\omega_x=0$) separately, where an
entirely analytic variational solution exists (Section \ref{s_waveguide}).
Subsequently, in Section \ref{s_3dcompare}, we again compare the ansatz
solutions to high-accuracy numerics.

\subsection{Variational solution: Gaussian ansatz\label{s_3dstart}}
We first consider an ansatz composed of Gaussian axial and radial
profiles. We phrase this as 
\begin{equation}
\psi(\mathbf{r})= \frac{\kappa^{1/2} \gamma^{3/4} k_\textrm{G}}{\pi^{3/4} \ell_{\textrm{G}}^{1/2}}e^{-\kappa \gamma k_{\textrm{G}}^2 (y^2 + z^2)/2}e^{-\gamma x^{2}/2 \ell_\textrm{G}^2}.
\end{equation}
Here, the first variational parameter, $\ell_\textrm{G}$, quantifies the axial
length of the ansatz in analogy to the quasi-1D case. The \textit{reciprocal}
of the second variational parameter, $k_{\textrm{G}}^{-1}$, quantifies the
radial length of the ansatz. In the trap-dominated limit ($\gamma \rightarrow
\infty$) both these lengths approach unity
($\{\ell_{\textrm{G}},k_\textrm{G}\}\rightarrow 1$).  Substitution of this
ansatz into Eq.\ (\ref{Eq:3DFunctional}) yields (using identities from Appendix
\ref{appendix_integrals}) 
\begin{equation}
H_{3\mathrm{D}}(\ell_{\textrm{G}},k_\textrm{G}) = 
\frac{\gamma}{4} \left( \ell_\textrm{G}^2 + \frac{1}{\ell_\textrm{G}^2} - \frac{2 k_\textrm{G}^2}{(2\pi\gamma)^{1/2} \ell_\textrm{G}}
+2 \kappa k_\textrm{G}^2 + \frac{2 \kappa}{k_\textrm{G}^2} \right)
\end{equation}
Setting the partial derivatives with respect to both $\ell_{\textrm{G}}$ and
$k_\textrm{G}$ equal to zero, we deduce that $\ell_\textrm{G}$ must solve the
quartic equation
\begin{equation}
\ell_{\textrm{G}}^{4} + \frac{k_\textrm{G}^2 \ell_{\textrm{G}}}{(2\pi\gamma)^{1/2}} -1 = 0,
\label{Eq:GaussGaussQuartic}
\end{equation}
and that $k_{\textrm{G}}$ must solve
\begin{equation}
k_\textrm{G} = \left(\frac{(2\pi\gamma)^{1/2}\kappa \ell_\textrm{G}}{(2\pi\gamma)^{1/2}\kappa \ell_\textrm{G} - 1} \right)^{1/4}.
\label{Eq:sSolutionGauss}
\end{equation}

From Eq.\ (\ref{Eq:sSolutionGauss}) it follows that we must have
$\ell_{\textrm{G}}> 1/(2\pi\gamma)^{1/2}\kappa$ to obtain a physically
reasonable solution, i.e., a real, positive value of $k_\textrm{G}$, consistent
with our initial ansatz.
For a given such value of $k_\textrm{G}$, Eq.\ (\ref{Eq:GaussGaussQuartic}) is
solved (see solution in Appendix \ref{appendix_quartic})
by
\begin{equation}
\ell_{\textrm{G}} = \frac{\left[\chi\left(\gamma k_\textrm{G}^{-4}\right)\right]^{1/2}k_\textrm{G}^{2/3}}{2^{4/3}(\pi\gamma)^{1/6}}
\left\{
\left[
\left(
\frac{2}{\chi\left(\gamma k_{\textrm{G}}^{-4}\right)}
\right)^{3/2}
-1
\right]^{1/2}
-1
\right\},
\label{Eq:EllValueGG}
\end{equation}
with $\chi$ defined as in Eq.\ (\ref{equation_chi_def}).

\subsection{Analysis of Gaussian ansatz solution\label{subsec_gg_analysis}}
\begin{figure*}
\includegraphics[width=6.75in]{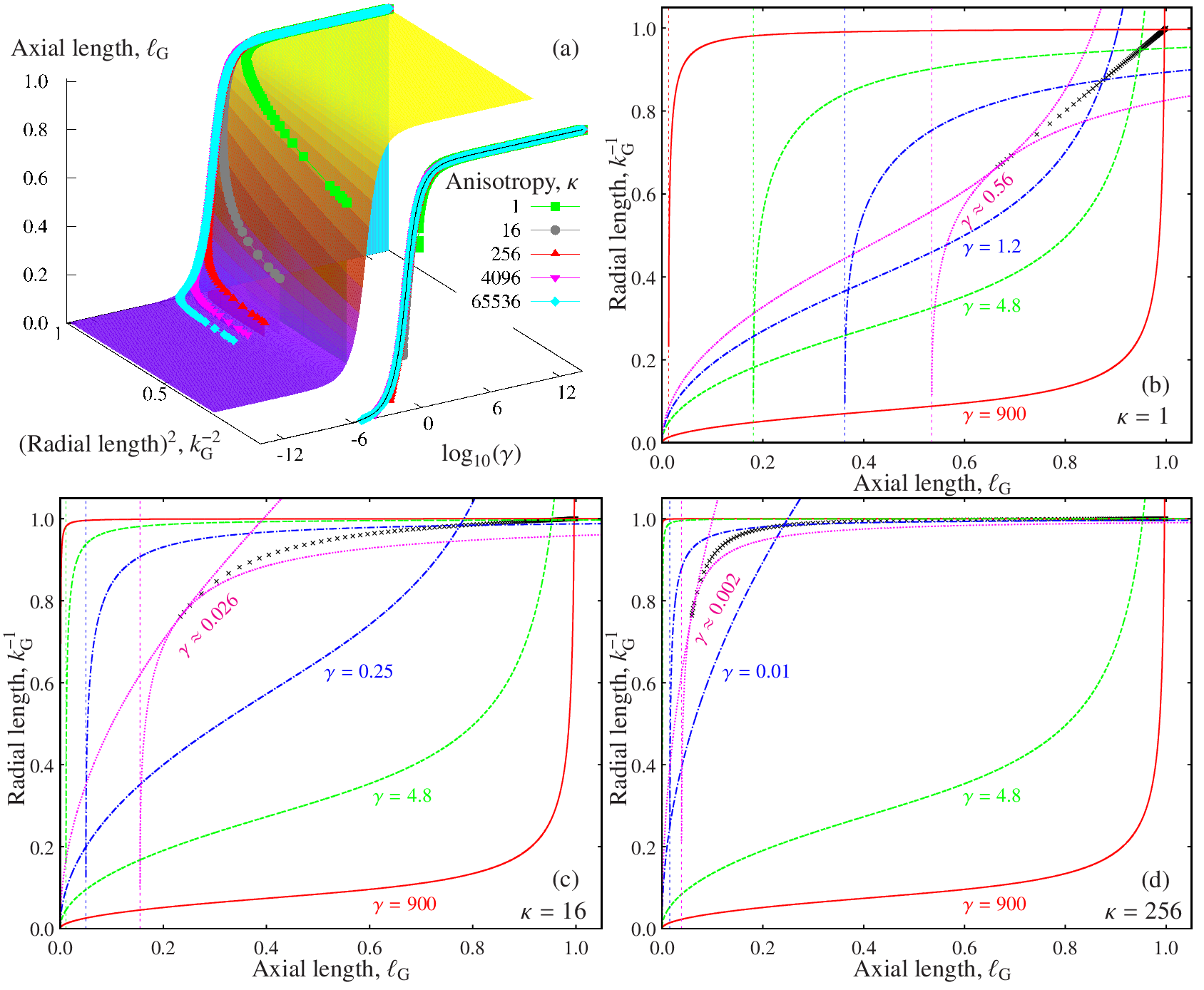}
\caption{Energy-minimizing variational parameters for the 3D GPE using a
Gaussian ansatz: (a) axial length $\ell_\textrm{G}$ as a function of the radial
length $k_\textrm{G}^{-1}$ and the parameter $\gamma$ [Eq.\
\ref{Eq:EllValueGG}]. Lines show the simultaneous solutions of equations
(\ref{Eq:sSolutionGauss}) and (\ref{Eq:sIntersectGauss}) for the axial length
$\ell_\textrm{G}$ and radial length $k_\textrm{G}^{-1}$, for different
anisotropies $\kappa$ and values of $\gamma$.  Projections of these solutions
on the $\gamma$--$\ell_\textrm{G}$ plane are also shown; here the black line
indicates the quasi-1D result [from figure \ref{figure_energies_1d}(a)]. (b--d)
Illustration of the intersections of equations (\ref{Eq:sSolutionGauss}) [lines
with vertical asymptote $\ell_\mathrm{G}=1/(2\pi\gamma)^{1/2}\kappa$ shown with
fine dashes] and (\ref{Eq:sIntersectGauss}) for various $\kappa$: the
higher-$\ell_\textrm{G}$ intersection, which corresponds to a physical solution
for the axial length $\ell_\textrm{G}$ and radial length $k_\textrm{G}^{-1}$,
can be found using a ``staircase'' method starting from $k_\textrm{G}=1$. The
numerical solutions obtained this way, and shown by points in (a), are shown by
crosses in (b--d).  The lowest values of $\gamma$ plotted in (b--d) are the
lowest for which a self-consistent Gaussian ansatz solution is found.}
\label{figure_gauss_3d}
\end{figure*}

Contrary to the quasi-1D limit, minimization of the variational energy in 3D
requires simultaneous solution of two equations for the radial length,
$k_\textrm{G}^{-1}$, and the axial length, $\ell_\textrm{G}$. These equations
are, respectively, Eq.\
(\ref{Eq:sSolutionGauss}) and [rearranged from Eq.\
(\ref{Eq:GaussGaussQuartic})]
\begin{equation}
k_\textrm{G} = \left[\frac{(2\pi\gamma)^{1/2} }{\ell_\textrm{G}} \left(1-\ell_\textrm{G}^4 \right) \right]^{1/2}.
\label{Eq:sIntersectGauss}
\end{equation}
These equations dictate that physical solutions must have 
\begin{equation}
\frac{1}{(2\pi\gamma)^{1/2}\kappa} < \ell_{\textrm{G}} < 1,
\label{ineq_gauss}
\end{equation}
and hence that $\gamma > 1/2\pi\kappa^2$ must be satisfied in order for
physical solutions to exist.

Where solutions exist, they must be found numerically.  However, a very
practical method of numerical solution follows from the shape of the
$\ell_\textrm{G}$ surface defined by Eq.\ (\ref{Eq:EllValueGG}), and shown in
Fig. \ref{figure_gauss_3d}(a), which is a decreasing function of $k_\textrm{G}$
for all (real, positive) $\gamma$.  The method can be considered graphically,
in terms of locating the intersection(s) of Eq.\ (\ref{Eq:sSolutionGauss}) and
Eq.\ (\ref{Eq:sIntersectGauss}).  These curves are shown, for various $\kappa$,
in Fig.  \ref{figure_gauss_3d}(b--d), along with the lower bound from
inequality (\ref{ineq_gauss}).  Below a $\kappa$-dependent threshold value of
$\gamma$ the curves fail to intersect, indicating instability of the BEC
to collapse. At the threshold value [dotted curves in Fig.\
\ref{figure_gauss_3d}(b--d)] there is exactly one intersection, and above the
threshold value [other curves in Fig.\ \ref{figure_gauss_3d}(b--d)] there are
two intersections. In the latter case the higher-$\ell_\textrm{G}$
intersection, which smoothly deforms to the limiting case
$\{\ell_{\textrm{G}},k_\textrm{G}\}\rightarrow 1$ as $\gamma \rightarrow
\infty$, represents the physical, minimal-energy variational solution. This
solution can be located using a simple ``staircase'' method: substituting a
trial value $\bar{k}_\textrm{G}$, satisfying $1 \leq \bar{k}_\textrm{G} <
k_\textrm{G}$, into Eq. \ref{Eq:EllValueGG} produces a trial value,
$\bar{\ell}_\textrm{G}$, satisfying $\ell_\textrm{G} < \bar{\ell}_\textrm{G}
\leq 1$, and subsequently substituting this trial value into Eq.
\ref{Eq:sSolutionGauss} produces an iterated trial value,
$\bar{k}_\textrm{G}^\prime$, satisfying $\bar{k}_\textrm{G} <
\bar{k}_\textrm{G}^\prime < k_\textrm{G}$. Thus, beginning with
$\bar{k}_\textrm{G} =1$, iteration of this process converges the trial values
to the true $k_\textrm{G}$ and $\ell_\textrm{G}$.

The physical solutions to equations (\ref{Eq:GaussGaussQuartic}) and
(\ref{Eq:sSolutionGauss}) for different anisotropies $\kappa$ are shown on the
$\ell_\textrm{G}$ surface, and projected into the $\ell_\textrm{G}$--$\gamma$
plane, in Fig.\ \ref{figure_gauss_3d}(a). These solutions are also shown as
black crosses in the $\ell_\textrm{G}$--$k_\textrm{G}$ plane in Figs.\
\ref{figure_gauss_3d}(b--d), where they form a line connecting the
physical-solution intersections of Eq.\ (\ref{Eq:sSolutionGauss}) and Eq.\
(\ref{Eq:sIntersectGauss}) for the various $\gamma$ shown. In Fig.\
\ref{figure_gauss_3d}(a) the collapse instability is manifest as a rapid rise
in $k_\textrm{G}$ --- corresponding to a decrease in radial extent --- and fall
in $\ell_\textrm{G}$ --- corresponding to a decrease in axial extent --- just
above a $\kappa$-dependent threshold value of $\gamma$. There are no
self-consistent solutions for these quantities below this collapse threshold.
For increasing anisotropies $\kappa$, this collapse threshold occurs at lower
values of $\gamma$. For the highest two values of $\kappa$ considered the
collapse threshold lies in the regime where $\ell_\textrm{G}$ is already
approaching $0$; our analysis of the Gaussian ansatz in the quasi-1D limit
indicates that the 3D Gaussian ansatz will be a poor approximation to the true
solution in this regime.  Importantly, for $\gamma$ above the collapse
threshold the projected curves for each anisotropy agree well with the Gaussian
ansatz in the quasi-1D GPE, suggesting that the Gaussian ansatz gives a good
approximation to the true solution here.

\subsection{Variational solution: soliton ansatz}
\begin{figure*}
\includegraphics[width=6.75in]{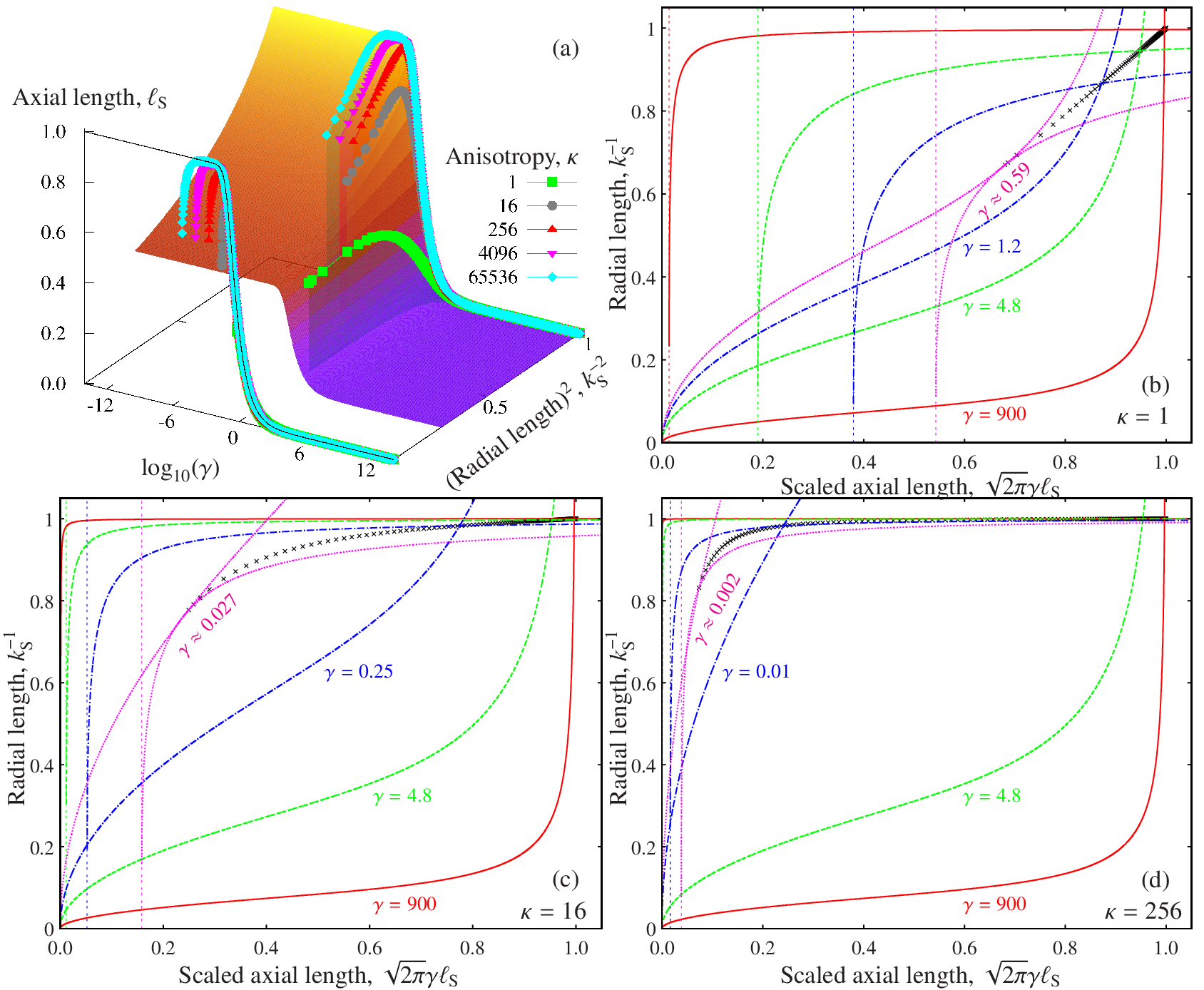}
\caption{Energy-minimizing variational parameters for the 3D GPE using a
soliton ansatz: (a) axial length $\ell_\textrm{S}$ as a function of the radial
length $k_\textrm{S}^{-1}$ and the parameter $\gamma$ [Eq.\
\ref{Eq:EllValueGS}]. Lines show the simultaneous solutions of equations
(\ref{Eq:sSolutionSech2}) and (\ref{Eq:sIntersect2}) for the axial length
$\ell_\textrm{S}$ and radial length $k_\textrm{S}^{-1}$, for different
anisotropies $\kappa$ and values of $\gamma$.  Projections of these solutions
on the $\gamma$--$\ell_\textrm{S}$ plane are also shown; here the black line
indicates the quasi-1D result [from figure \ref{figure_energies_1d}(a)]. (b--d)
Illustration of the intersections of equations (\ref{Eq:sSolutionSech2}) [lines
with vertical asymptote $\ell_\mathrm{S}=(\pi/3)/(2\pi\gamma)^{1/2}\kappa$
shown with fine dashes] and (\ref{Eq:sIntersect2}) for various $\kappa$: the
higher-$\ell_\textrm{S}$ intersection, which corresponds to a physical solution
for the axial length $\ell_\textrm{S}$ and radial length $k_\textrm{S}^{-1}$,
can be found using a ``staircase'' method starting from $k_\textrm{S}=1$. The
numerical solutions obtained this way, and shown by points in (a), are shown by
crosses in (b--d).  The lowest values of $\gamma$ plotted in (b--d) are the
lowest for which a self-consistent soliton ansatz solution is found.}
\label{figure_sech_3d}
\end{figure*}

Secondly, we consider a soliton ansatz composed of a axial sech profile and a
radial Gaussian profile. We phrase this as 
\begin{equation}
\psi(\mathbf{r})=\frac{\gamma^{1/2}\kappa^{1/2} k_\textrm{S}}{(2\pi\ell_{\textrm{S}})^{1/2}}e^{-\kappa \gamma k_\textrm{S}^2 (y^{2}+z^{2})/2}\mbox{sech}(x/2\ell_{\textrm{S}}).
\end{equation}
As with the 3D Gaussian ansatz, the first variational parameter,
$\ell_\textrm{G}$, quantifies the axial length of the ansatz and the
\textit{reciprocal} of the second variational parameter, $k_{\textrm{G}}^{-1}$,
quantifies its radial length. In the quasi-1D limit both lengths consequently
approach unity ($\{\ell_{\textrm{G}},k_\textrm{G}\}\rightarrow 1$).
Substituting this ansatz into Eq.\ (\ref{Eq:3DFunctional}) yields (using
identities from Appendix \ref{appendix_integrals})
\begin{multline}
H_{3\mathrm{D}}(\ell_{\textrm{S}},k_\textrm{S}) = \frac{\pi^2\gamma^2}{6} \left(\ell_\textrm{S}^2 + \frac{1}{4\pi^2\gamma^2\ell_\textrm{S}^2} - \frac{k_\textrm{S}^2}{2\pi^2\gamma^2\ell_\textrm{S}}\right.
\\ +\left.\frac{3\kappa k_\textrm{S}^2}{\pi^2\gamma} + \frac{3\kappa}{\pi^2\gamma k_\textrm{S}^2}\right).
\end{multline}
Once again, setting partial derivatives with respect to both
$\ell_{\textrm{S}}$ and $k_\textrm{S}$ equal to zero allows us to deduce that:
\begin{equation}
\ell_{\textrm{S}}^{4}
+ \frac{k_\textrm{S}^2\ell_{\textrm{S}}}{4\pi^2\gamma^{2}}
- \frac{1}{4\pi^2\gamma^{2}} = 0,
\label{Eq:GaussSechQuartic}
\end{equation}
and that $k_{\textrm{S}}$ must solve
\begin{equation}
k_\textrm{S} = \left(\frac{6\kappa \gamma\ell_\textrm{S}}{6\kappa \gamma\ell_\textrm{S} - 1} \right)^{1/4}.
\label{Eq:sSolutionSech}
\end{equation}

From Eq.\ (\ref{Eq:sSolutionSech}) it follows that we must have
$\ell_{\textrm{S}}> 1/6\kappa\gamma$ to obtain a physically reasonable
solution, i.e., a real, positive value of $k_\textrm{S}$, consistent with our
initial ansatz.  For a given such value of $k_\textrm{S}$, Eq.\
(\ref{Eq:GaussSechQuartic}) is solved (see solution in Appendix
\ref{appendix_quartic}) by
\begin{equation}
\ell_{\textrm{S}} = \frac{\left[\chi\left(\gamma k_{\textrm{S}}^{-4}\right)\right]^{1/2}k_\textrm{S}^{2/3}}{2^{11/6}(\pi\gamma)^{2/3}}
\left\{
\left[
\left(
\frac{2}{\chi\left(\gamma k_{\textrm{S}}^{-4}\right)}
\right)^{3/2}
-1
\right]^{1/2}
-1
\right\},
\label{Eq:EllValueGS}
\end{equation}
with $\chi$ defined as in Eq.\ (\ref{equation_chi_def}).

\subsection{Analysis of soliton ansatz solution\label{s_3dend}}
As in the case of the Gaussian ansatz, minimization of the variational energy
in 3D requires the simultaneous solution of equations for the radial length
$k_\textrm{S}^{-1}$ and the axial length $\ell_\textrm{S}$. These equations
are, respectively, Eq.\ (\ref{Eq:sSolutionSech}) and [rearranged from Eq.\
(\ref{Eq:GaussSechQuartic})]
\begin{equation}
k_\textrm{S} = \left[\frac{1}{\ell_\textrm{S}} \left(1-4\pi^2\gamma^2\ell_\textrm{S}^4 \right) \right]^{1/2}
\label{Eq:sIntersect}
\end{equation}
These equations dictate that physical solutions must have 
\begin{equation}
\frac{1}{6\kappa\gamma} < \ell_{\textrm{S}} < \frac{1}{(2\pi\gamma)^{1/2}}
\label{ineq_sech}
\end{equation}
and hence that $\gamma > (\pi/3)^2/2\pi\kappa^2$ must be satisfied in order for
physical solutions to exist. These equations and constraints can be further
simplified by casting them in terms of $\ell_\textrm{S}^\prime =
(2\pi\gamma)^{1/2}\ell_\textrm{S}$; this yields two equations,
\begin{equation}
k_\textrm{S} = \left(\frac{(2\pi\gamma)^{1/2}\kappa \ell_\textrm{S}^\prime}{(2\pi\gamma)^{1/2}\kappa \ell_\textrm{S}^\prime - \pi/3} \right)^{1/4}
\label{Eq:sSolutionSech2}
\end{equation}
and
\begin{equation}
k_\textrm{S} = \left[\frac{(2\pi\gamma)^{1/2}}{\ell_\textrm{S}^\prime} \left(1-{\ell_\textrm{S}^\prime}^4 \right) \right]^{1/2},
\label{Eq:sIntersect2}
\end{equation}
and an inequality,
\begin{equation}
\frac{\pi/3}{(2\pi\gamma)^{1/2}\kappa} < \ell_{\textrm{S}}^\prime < 1,
\label{ineq_sech2}
\end{equation}
which are extremely similar to those encountered in the case of the Gaussian
ansatz. The numerical solution of these equations for the physical solution,
which can only exist when $\gamma > (\pi/3)^2/2\pi\kappa^2$, follows the same
procedure as used for the Gaussian ansatz.

Variational-energy-minimizing solutions to the soliton ansatz equations for
different anisotropies $\kappa$ are shown in Fig.\ \ref{figure_sech_3d}; these
are shown superimposed on the $\ell_\textrm{S}$ surface and projected into the
$\ell_\textrm{S}$--$\gamma$ plane in Fig.\ \ref{figure_sech_3d}(a), and
alongside equations (\ref{Eq:GaussSechQuartic}) and (\ref{Eq:sSolutionSech})
and inequality (\ref{ineq_sech2}) in Fig.\ \ref{figure_sech_3d}(b--d). The
collapse instability is even more evident in the soliton ansatz than in the
Gaussian ansatz, since it occurs in a region with a larger background value of
$\ell_\textrm{S}$. Once again, the collapse is manifest as a rapid rise in
$k_\textrm{S}$ and drop in $\ell_\textrm{S}$ --- corresponding to both axial
and radial contraction of the solution--- immediately prior to a
$\kappa$-dependent threshold value of $\gamma$. Below the threshold, no
self-consistent solutions exist.  For increasing anisotropies $\kappa$, this
collapse threshold again occurs at lower values of $\gamma$.  In contrast to
the case of the Gaussian ansatz, however, the collapse instability precludes
solutions in exactly the limit where one expects the soliton ansatz to be
accurate ($\gamma \rightarrow 0$). This property of the collapse instability
severely restricts the possibility of observing highly bright-soliton-like
ground states in 3D.  The solution curves in Fig.\ \ref{figure_sech_3d}(a)
illustrate that this effect is worst for low trap anisotropies $\kappa$, but is
to some extent mitigated for higher $\kappa$. However, a full comparison with
numerically exact solutions is necessary to quantify these effects; we
undertake such a comparison in Section \ref{s_3dcompare}.

\subsection{Variational solution: waveguide configuration\label{s_waveguide}}
In broad experimental terms, the collapse instability sets a maximum value for
the ratio of interaction strength to trap strength (equivalent to a minimum
value of $\gamma$) which increases (and hence the minimum value of $\gamma$
decreases) with the trap anisotropy $\kappa$. In the context of atomic BEC
experiments one would typically think of controlling the interaction--trap
strength ratio by varying either $|a_s|$ or $N$ while holding $\omega_r$ and
$\omega_x$ constant; in this situation the collapse instability places a
trap-anisotropy-dependent upper limit on the product $|a_s|N$.  However, the
minimum value of $\gamma$ does not increase without limit in the trap
anisotropy $\kappa$: In an experiment one can, in principle, remove all axial
trapping to create a waveguide-like configuration; in this case $\omega_x=0$
and the trap anisotropy $\kappa \rightarrow \infty$, while the parameter
$\gamma \rightarrow 0$.  In this limit a reparametrization is necessary, and
only needs to be performed for the soliton ansatz, which is clearly more
appropriate in this context.

\begin{figure}
\includegraphics[width=3.375in]{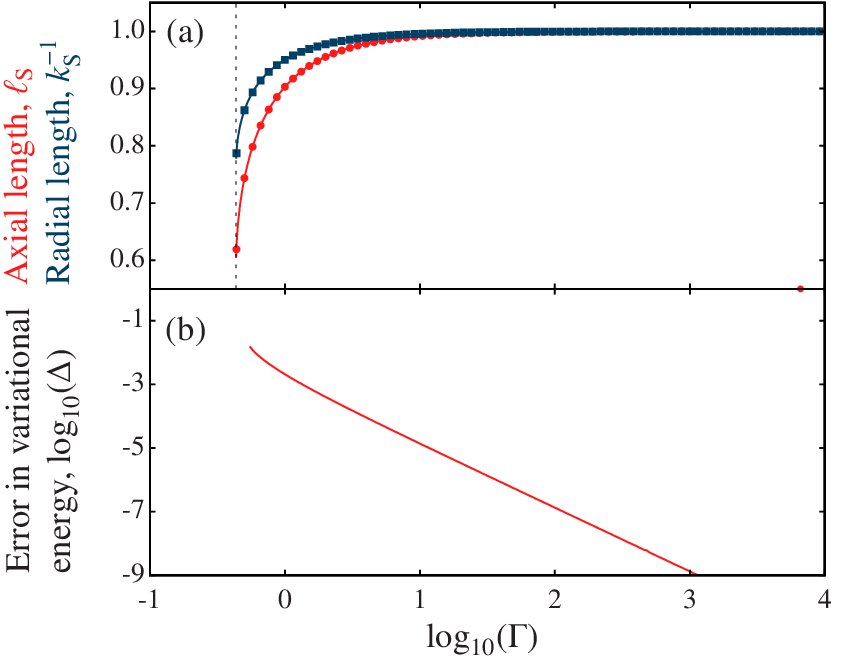}
\caption{Comparison of 3D variational and numerical solutions in a waveguide
configuration ($\omega_x=0$: (a) Energy-minimizing axial length
$\ell_\textrm{S}$ and radial length $k_\textrm{S}^{-1}$ for the soliton ansatz.
Solutions, given by Eq.\ (\ref{Eq:WaveSolution}), exist for all $\Gamma =
\kappa\gamma > 3^{1/2}/4$. (b) Relative error in the minimum variational energy
of the soliton ansatz, $\Delta = (H_\textrm{3D} -
E_\textrm{3D})/E_\textrm{3D}$, where $E_\textrm{3D}$ is the numerically
determined ground state energy.}
\label{figure_sim_waveguide}
\end{figure}

Elimination of the axial trap eliminates one of the two free parameters of the
3D GPE [Eq.\ (\ref{Eq:Scaled3DGPE})]. The remaining free parameter is $\Gamma =
\gamma\kappa = (a_{r}/2|a_s|N)^2$, where $a_\mathrm{r} =
(\hbar/m\omega_r)^{1/2}$ is the radial harmonic oscillator length scale. The
soliton ansatz may be re-written in terms of $\Gamma$ as 
\begin{equation}
\psi(\mathbf{r})=\frac{\Gamma^{1/2} k_\textrm{S}}{(2\pi\ell_{\textrm{S}})^{1/2}}e^{-\Gamma k_\textrm{S}^2 (y^{2}+z^{2})/2}\mbox{sech}(x/2\ell_{\textrm{S}}).
\end{equation}
Substituting this into Eq.\ (\ref{Eq:3DFunctional}) with $\omega_x=0$ yields
(using identities from Appendix \ref{appendix_integrals}),
\begin{equation}
H_{3\mathrm{D}}(\ell_{\textrm{S}},k_\textrm{S}) = \left( \frac{1}{24\ell_\textrm{S}^2} - \frac{k_\textrm{S}^2}{12\ell_\textrm{S}}
+\frac{\Gamma k_\textrm{S}^2}{2} + \frac{\Gamma}{2 k_\textrm{S}^2}\right),
\end{equation}
from which we deduce that the energy-minimizing variational parameters satisfy
\begin{equation}
\ell_{\textrm{S}} = \frac{1}{k_\textrm{S}^2}
\label{Eq:WaveL}
\end{equation}
and
\begin{equation}
k_\textrm{S} = \left(\frac{6\Gamma\ell_\textrm{S}}{6\Gamma\ell_\textrm{S} - 1} \right)^{1/4}.
\label{Eq:WaveK}
\end{equation}
Contrary to the more general 3D case, an analytic simultaneous solution of Eq.\
(\ref{Eq:WaveL}) and Eq.\ (\ref{Eq:WaveK}) exists when $\ell_\textrm{S}$
satisfies the depressed cubic equation
\begin{equation}
\ell_\textrm{S}^3 - \ell_\textrm{S} + \frac{1}{6\Gamma} = 0.
\label{Eq:WaveMaster}
\end{equation}
Using the general solution for a depressed cubic equation from Appendix
\ref{appendix_quartic}, one finds that the physical root (with real, positive
$\ell_\textrm{S}$ satisfying the limit $\ell_\textrm{S}\rightarrow 1$ as
$\Gamma \rightarrow \infty$) is given by
\begin{multline}
\ell_\textrm{S} = \left[-\frac{1}{12\Gamma} + \frac{1}{3^{3/2}\Gamma}\left(\frac{3}{16}-\Gamma^2 \right)^{1/2} \right]^{1/3} \\
+ \left[-\frac{1}{12\Gamma} - \frac{1}{3^{3/2}\Gamma}\left(\frac{3}{16}-\Gamma^2 \right)^{1/2} \right]^{1/3}.
\label{Eq:WaveSolution}
\end{multline}
Consequently, solutions only exist for $\Gamma > 3^{1/2}/4$, as shown in Fig.\
\ref{figure_sim_waveguide}(a).

\subsection{Comparison to 3D numerical solutions\label{s_3dcompare}}
\begin{figure}
\includegraphics[width=3.375in]{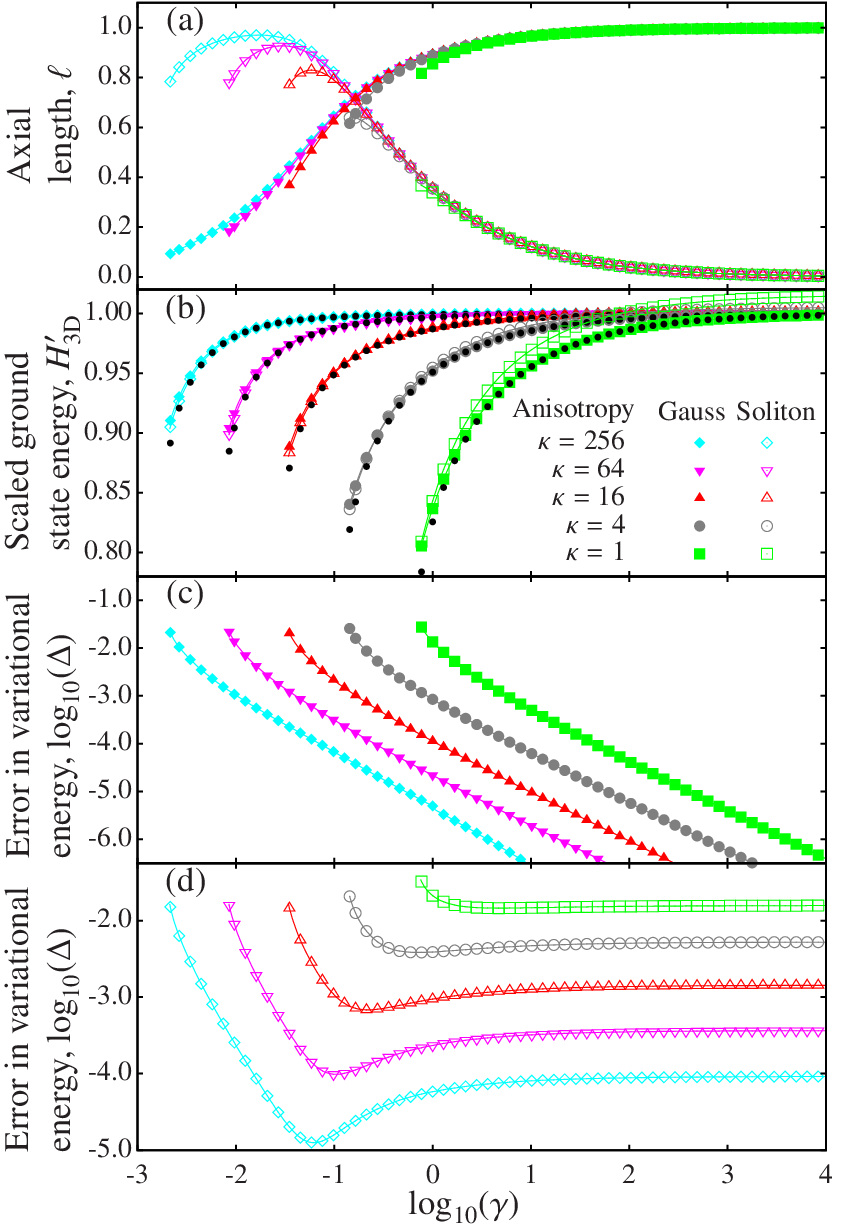}
\caption{Comparison of 3D variational and numerical solutions: (a)
Energy-minimizing axial lengths $\ell_\textrm{G}$ (Gaussian ansatz, solid
symbols) and $\ell_\textrm{S}$ (soliton ansatz, hollow symbols). (b) Scaled
variational energies $H_\textrm{3D}^\prime = \kappa
H_\textrm{3D}/\gamma(\kappa+1/2)$ (a similarly scaled ground state energy
$E_\textrm{3D}^\prime = \kappa E_\textrm{3D}/\gamma(\kappa+1/2)$ tends to $1$
in the limit $\gamma \rightarrow \infty$ for all anisotropies $\kappa$)
compared with the numerically calculated ground state energies $E_\textrm{3D}$
(black dots).  (c,d) Normalized relative error in the variational energy
$\Delta = (H_\textrm{3D} - E_\textrm{3D})/E_\textrm{3D}$ for the Gaussian (c)
and soliton (d) ansatzes. For clarity every 4th datum is marked by a symbol in
(a--d).}
\label{figure_energies_3d}
\end{figure}

The variational energy-minimizing axial lengths $\ell_\textrm{G}$ and
$\ell_\textrm{S}$ are shown as functions of $\gamma$ in Fig.\
\ref{figure_energies_3d}(a) for the general 3D case; for the waveguide limit
both axial and radial lengths $\ell_\textrm{S}$ and $k_\textrm{S}^{-1}$ are
shown as functions of $\Gamma$ in Fig.\ \ref{figure_sim_waveguide}(a). As in
the quasi-1D case, we quantitatively evaluate the accuracy of the ansatz
solutions for general $\gamma$ ($\Gamma$) by comparing the variational minimum
energy $H_\textrm{3D}$ with the numerically determined ground state energy
$E_\textrm{3D}$. We calculate $E_\textrm{3D}$ using a pseudospectral method in
a basis of optimally-scaled harmonic oscillator eigenstates; this is formed
from a tensor product of symmetric Gauss-Hermite functions (axial direction)
and generalized Laguerre functions (radial direction). The ansatz with the
lowest variational energy is used both to optimize the scaling of the basis
functions and as an initial estimate for the solution. Expanding the stationary
3D GPE in such a basis produces a system of nonlinear equations which are
solved iteratively using a modified Newton method. A similar method was used to
solve a similar cylindrically symmetric, stationary 3D GPE, with repulsive
interactions, in Ref. \cite{morgan_pra_2005}.

As in the quasi-1D case, we compare several quantities between the ansatz and
numerical solutions.  Fig.\ \ref{figure_energies_3d}(b) shows the
\textit{scaled} energy $H_\textrm{3D}^\prime =
(H_\textrm{3D}/\gamma)/(1+1/2\kappa)$ in the general 3D case. This scaling is
such that $E_\textrm{3D}^\prime$ --- which is defined analogously to
$H_\textrm{3D}^\prime$ with respect to $E_\textrm{3D}$ --- tends to 1 as
$\gamma\rightarrow \infty$. Figs.\ \ref{figure_energies_3d}(c) and (d) show the
relative error in the variational minimum energy $\Delta = (H_\textrm{3D} -
E_\textrm{3D})/E_\textrm{3D}$ for the Gaussian and soliton ansatzes,
respectively. The same quantity $\Delta$ is shown for the waveguide limit in
Fig.\ \ref{figure_sim_waveguide}(b). All quantities shown in Figs.\
\ref{figure_energies_3d} and \ref{figure_sim_waveguide} are computed using
between 2000 and 12000 basis states ($\kappa$-dependent) and are insensitive to
a doubling of the number of basis states.

In the general 3D case, a close inspection of Fig.\
\ref{figure_energies_3d}(b--d) is necessary to reveal the overall relation
between the ansatz solutions and the numerically obtained ground state. In the
high-$\gamma$ limit Fig.\ \ref{figure_energies_3d}(b) shows that both the
Gaussian variational energies (solid symbols) and the ground state energy
$E_\textrm{3D}$ (black dots) approach 1 as $\gamma \rightarrow \infty$, whereas
the soliton ansatz energies (hollow symbols) tend to higher energies. This
corresponds to the actual ground state most closely matching the Gaussian
ansatz in this limit, as one would expect. Indeed, the relative error in
variational energy, $\Delta$, for the Gaussian ansatz [Fig.\
\ref{figure_energies_3d}(c)] continues to drop exponentially with $\gamma$ for
all anisotropies $\kappa$, making it possible to find regimes of $\gamma$ where
the Gaussian ansatz gives an excellent approximation to the true ground state.

In the opposite, low-$\gamma$ limit, collapse occurs at a $\kappa$-dependent
value of $\gamma$; this corresponds to the points in Fig.\
\ref{figure_energies_3d}(a--d) where solution curves abruptly cease. Prior to
collapse (at higher values of $\gamma$) the relation between the Gaussian
ansatz, the soliton ansatz, and the actual ground state is highly dependent on
the trap anisotropy $\kappa$ [Fig.\ \ref{figure_energies_3d}(b)]. In the case
of a spherically symmetric trap, where the anisotropy $\kappa=1$, the soliton
ansatz variational energy is \textit{never} closer to the true ground state
energy $E_\textrm{3D}$ than the Gaussian ansatz variational energy. A regime of
soliton-like ground states consequently cannot exist at this low anisotropy; as
the soliton ansatz is intrinsically asymmetric, this is to be expected.  For
higher anisotropies, the soliton ansatz energy is closer to $E_\textrm{3D}$
than the Gaussian ansatz energy in a small regime prior to collapse. Exactly
how soliton-like the ground state is in this regime can be quantitatively
assessed using the relative error $\Delta$. This is shown for the soliton
ansatz in [Fig.\ \ref{figure_energies_3d}(d)]. For each $\kappa$ the
``background'' value of $\Delta$ in the limit $\gamma \rightarrow \infty$ is
different; this effect is due to the decreasing size of the axial part of the
energy with respect to the radial part for increasing $\gamma$. In the
opposite, low-$\gamma$, limit $\Delta$ increases sharply close to the collapse
point as the ground state wavefunction rapidly contracts. The maximum extent to
which $\Delta$ decreases from its high-$\gamma$ limit, \textit{before} this
increase due to collapse-related contraction at low $\gamma$, quantifies how
soliton-like the ground state becomes in this regime. Even for the highest
anisotropy shown, $\kappa=256$, the regime of $\gamma$ over which $\Delta$
drops below its background value is rather narrow, and the actual drop in
$\Delta$ is only one order of magnitude. Compared to benchmark of Section
\ref{s_1dcompare}, this indicates that the true ground state remains
considerably deformed with respect to the soliton ansatz. The minimum error in
the soliton ansatz energy does, however, improve with increasing anisotropy
$\kappa$. Excellent agreement can be achieved in the waveguide limit ($\kappa
\rightarrow \infty$): Fig.\ \ref{figure_sim_waveguide} shows that excellent
agreement, with respect to the benchmark figure of Section \ref{s_1dcompare},
can be obtained for $\Gamma>10^{3/2}$.

\subsection{Discussion\label{s_discuss}}
A physical interpretation of the above results follows from considering two
conditions that must be satisfied in order to realize a soliton-like ground
state; (1) the radial profile should be ``frozen'' to a Gaussian, thus
realizing a quasi-1D limit; and (2) interactions should dominate over the axial
trapping. On first inspection these conditions seem mutually compatible, and
satisfiable simply by increasing the radial trap frequency $\omega_r$ with
other parameters held constant.  However, condition (1) can only be satisfied
if the maximum density remains low enough to avoid any deformation of the
radial profile due to the collapse instability.  Increasing $\omega_r$
leads to exactly such deformation, and ultimately to collapse, as it has the
secondary effect of strongly increasing the density. This strong increase in
density with $\omega_r$ is particular to the case of attractive interactions.
Increasing $\omega_r$ in a repulsively-interacting BEC likewise acts to
increase the density, but this increase is counteracted by the interactions;
these act to reduce the density, and cause the BEC to expand axially. In the
attractively-interacting case the response of the interactions is the opposite:
increasing $\omega_r$ leads to axial \textit{contraction} of the BEC.
Consequently condition (1) is far harder to satisfy for an
attractively-interacting BEC than a repulsively-interacting one. Responding to
this problem simply by reducing the interaction strength (either through
$|a_s|$ or $N$) leads to violation of condition (2). The nature of the problem
is made particularly clear by considering the waveguide limit: here condition
(2) is automatically satisfied ($\omega_x = 0$). This makes it possible to
achieve a highly soliton-like ground state by satisfying condition (1) alone.
However, such a ground state is achieved by \textit{lowering} the product
$\omega_r^{1/2}|a_s|N$, and thus by progressing towards the limit of extreme
diluteness.

This physical behavior of the system presents considerable challenges for
experiments aiming to realize a highly soliton-like ground state.  In essence,
the most desirable configuration is to have extremely high anisotropies
$\kappa$, while keeping $\omega_r$ as low as possible.  Realizing such a
configuration through extremely low, or zero, axial trap frequencies $\omega_x$
is problematic: such frequencies are hard to set precisely experimentally as
they require a very smooth potential to be generated, potentially over a
considerable length. Furthermore, in the case $\omega_x=0$ the mean-field
approximation ceases to be valid for an attractively-interacting BEC; the true
wavefunction should be translationally invariant in this case, but the
mean-field solution breaks this symmetry \cite{ lai_haus_pra_1989_b}.  Even for
very low but non-zero $\omega_x$ the mean-field approximation can lose validity
due to the extreme diluteness of the BEC, and the energy gap from the ground
state to states with excited axial modes can become low enough to cause
significant population of the excited states at experimentally feasible
temperatures.

It is informative to consider the parameters used in bright solitary wave
experiments to date \cite{khaykovich_etal_science_2002,
strecker_etal_nature_2002, cornish_etal_prl_2006}. None of these aimed to
realize highly soliton-like ground states in the sense considered here.
However, they nonetheless indicate regimes which have proved to be
experimentally accessible and offer a guide to future possibilities. All have
operated outside the regime of highly soliton-like ground states; direct
comparison of the experiments of Refs.  \cite{strecker_etal_nature_2002} and
\cite{cornish_etal_prl_2006} with our results reveals that $\kappa$ is too
small in these experiments ($\kappa \approx 11$ and $\kappa \approx 3$
respectively) to achieve a highly soliton-like ground state. The experiment of
Ref.  \cite{khaykovich_etal_science_2002} featured an expulsive axial
potential, which does not yield a value of $\kappa$ suitable for direct
comparison with our results. However, it is possible to assume the waveguide
limit $\omega_x = 0$ in each experiment and compare the values of $\Gamma$ with
our results: in each case $\Gamma \lesssim 1$, outside the regime of highly
soliton-like ground states. Thus, experiments with weaker traps and lower
densities than previously realized with attractive condensates appear to be
necessary in order to achieve a highly soliton-like ground state.

\section{Conclusions\label{s_conclusions}}
In this paper we considered attractively-interacting atomic BECs in
cylindrically symmetric, prolate harmonic traps, and introduced variational
ansatzes, based on Gaussian and bright-soliton profiles, for the GPE ground
state. We compared new, analytic variational solutions based on these ansatzes
with highly accurate numerical solutions of the GPE over an extensive parameter
space, and hence determined how soliton-like the ground state is. Initially
assuming the quasi-1D limit to be valid, we showed that the true solution to
the GPE is (not) soliton like when interactions do (not) dominate over the trap
strength. In 3D, this picture is complicated by the collapse instability; in
the regime where all trap strengths dominate over the interactions a Gaussian
variational ansatz gives an excellent approximation to the true, and
non-soliton-like ground state.  In contrast to the quasi-1D limit, however, we
have shown that the regime in which the ground state is truly soliton-like
(well approximated by a soliton variational ansatz) is either non-existent, or
highly restricted, depending on the trap anisotropy.  For low anisotropies,
as one raises the strength of the interactions such that they approach and
exceed the strength of the axial trap the true ground state ceases to be
well-described by a Gaussian variational ansatz, but does not become
well-described by a soliton variational ansatz before the interaction strength
also exceeds the radial trapping strength, leading to collapse. Only by raising
the anisotropy significantly can one open a parameter window in which the
true ground-state becomes soliton-like before the interaction strength is
sufficient to cause collapse.

Our results describe the nature of the ground state over a wide parameter
regime, and offer a straightforward, accurate approximation to the full 3D GPE
solution in many cases. Our results are particularly relevant for experiments
using attractively-interacting condensates as they identify the potentially
challenging parameter regime required to observe a truly soliton-like ground
state, which would be an advantageous regime for experiments seeking to explore
and exploit beyond-mean field effects such as a macroscopic superposition of
bright solitons.  Given that previous studies have shown that the dynamics and
collisions of bright solitary waves can be soliton-like over a much wider
parameter regime than our approach reveals the ground state to be, extending
the variational approach used here to dynamical situations is an interesting
direction for future work.

\acknowledgments{
We thank S. L. Cornish, D. I. H. Holdaway, H. Salman and C. Weiss for
discussions, and the UK EPSRC (Grant No. EP/G056781/1), the Jack Dodd Centre
(S.A.G.) and Durham University (T.P.B.) for support.}

\begin{appendix}

\section{Useful integrals\label{appendix_integrals}}
Considering a Gaussian ansatz to be proportional to $e^{-k^{2}x^{2}}$, for completeness we reprise the following sequence of well-known integral identities, all of which are necessary to determine the corresponding variational energy functional:
\begin{gather}
\int_{-\infty}^{\infty} dx e^{-2k^{2}x^{2}} = \frac{\sqrt{\pi/2}}{k} \Rightarrow
\int_{-\infty}^{\infty} dx e^{-4k^{2}x^{2}} = \frac{\sqrt{\pi}}{2k},
\label{Eq:GaussID1}
\\
\int_{-\infty}^{\infty} dx x^{2}e^{-2k^{2}x^{2}} = -\frac{1}{4k}\frac{\partial}{\partial k}\int_{-\infty}^{\infty} dx e^{-2k^{2}x^{2}} = \frac{\sqrt{\pi/2}}{4k^{3}},
\label{Eq:GaussID2}
\\
\int_{-\infty}^{\infty} dx \left(\frac{\partial}{\partial x}e^{-k^{2}x^{2}}\right)^{2} =4k^{4}\int_{-\infty}^{\infty} dx x^{2} e^{-k^{2}x^{2}} = k\sqrt{\pi/2}.
\label{Eq:GaussID3}
\end{gather}

Comparable integral identities exist when considering an ansatz proportional to $\mbox{sech} (k x)$. Thus:
\begin{gather}\int_{-\infty}^{\infty} dx \mbox{sech}^{2}(k x) 
= \left[ \frac{\mbox{tanh}(k x)}{k}\right]_{-\infty}^{\infty} 
= \frac{2}{k}, 
\label{Eq:SechId1}\\
\int_{-\infty}^{\infty} dx \mbox{sech}^{4}(k x) = 
\left[
\frac{\{\mbox{sech}^{2}(k x) + 2\} \mbox{tanh}(k x)}{3k}
\right]_{-\infty}^{\infty} =
\frac{4}{3k}, 
\label{Eq:SechId2}
\\
\begin{split}
\int_{-\infty}^{\infty} dx \left[ \frac{\partial}{\partial x}
\mbox{sech}(k x) \right]^{2} = &
k^{2} \int_{-\infty}^{\infty} dx 
\mbox{tanh}^{2}(k x)\mbox{sech}^{2}(x)
\\
=& 
\frac{k}{3}\left[
\mbox{tanh}(k x)
\right]_{-\infty}^{\infty}
=\frac{2k}{3},
\end{split}
\label{Eq:SechId3}
\end{gather}
all of which are necessary to determine the energy of a standard bright soliton solution to the nonlinear Schr\"{o}dinger equation.  However, we also require a contribution arising from the existence of an external harmonic confining potential. Hence, we determine
\begin{equation}
\begin{split}
\int_{-\infty}^{\infty} dx x^{2}\mbox{sech}^{2}(k x) =& 2\int_{0}^{\infty} dx x^{2}\mbox{sech}^{2}(k x)\\
= &\frac{2}{k^{3}}
\Bigl[
\mbox{Li}_{2}\left(-e^{-2k x}\right) 
+ k x \Bigl\{
k x \mbox{tanh} (k x) 
\\ &
-k x
 - 2\ln\left(1+e^{-2k x}\right)
\Bigr\}\Bigr]_{0}^{\infty}
\\
=& \frac{2}{k^{3}}\left[
\mbox{Li}_{2}(0) - \mbox{Li}_{2}(-1)
\right]  \\ 
=& \frac{2}{k^{3}}\eta(2) = \frac{\pi^{2}}{6k^{3}},
\end{split}
\label{Eq:SechId4}
\end{equation}
where $\mbox{Li}_{y}(x) \equiv \sum_{n=1}^{\infty}x^{n}/n^{y}$ is a polylogarithm, and $-\mbox{Li}_{y}(-1) = \eta(y)$, the Dirichlet $\eta$ function, with $\eta(2) = \pi^{2}/12$.  

\section{Solution to the quartic equations\label{appendix_quartic}}

We require a general solution to a quartic in $\ell$ of the form
\begin{equation}
\ell^{4} + b\ell - c = 0,
\label{Eq:BaseQuartic}
\end{equation}
where $b$ and $c$ are positive real constants, and $\ell$ must also take positive real values to be physically meaningful.
This can be rephrased as the product of two quadratics in $\ell$:
\begin{equation}
\left[
\ell^{2} + \alpha \ell + \frac{1}{2}
\left(
\alpha^{2} - \frac{b}{\alpha}
\right)
\right]
\left[
\ell^{2} - \alpha \ell + \frac{1}{2}
\left(
\alpha^{2} + \frac{b}{\alpha}
\right)
\right]
= 0,
\label{Eq:QuadraticProduct}
\end{equation}
so long as
$
(
b^{2}/\alpha^{2}-\alpha^{4} 
)/4= c
$.  Hence, $\alpha$, which remains to be determined, must solve $\alpha^{6} + 4c\alpha^{2} - b^{2}$. 

Defining $\xi = \alpha^{2}$, the problem of determining $\alpha$ reduces to finding values of $\xi$  to solve the depressed cubic equation
\begin{equation}
\xi^{3} + 4c\xi - b^{2} = 0.
\label{Eq:AssociatedCubic}
\end{equation}
Defining 
\begin{equation}
A= \sqrt[3]{\frac{b^{2}}{2}+ \sqrt{\frac{b^{4}}{4}+\frac{64 c^{3}}{27}}},
\quad
B= \sqrt[3]{\frac{b^{2}}{2}- \sqrt{\frac{b^{4}}{4}+\frac{64 c^{3}}{27}}},
\end{equation}
the three roots of Eq.\ (\ref{Eq:AssociatedCubic}) are given by: 
\begin{align}
\xi_{1} =& A+B, \\
\xi_{2} =& -(A+B)/2 + i\sqrt{3}(A-B)/2, \\ 
\xi_{3} =& -(A+B)/2 - i\sqrt{3}(A-B)/2. 
\end{align} 
Any one of these will solve Eq.\ (\ref{Eq:AssociatedCubic}), however we choose $\xi_{1}$; as $b$ and $c$ are assumed positive real, $\xi_{1}$ is also conveniently guaranteed positive real. 

Substituting in $\alpha = \sqrt{\xi_{1}}$, we can apply the quadratic formula to both the factors (enclosed in square brackets) on the left hand side of Eq.\ (\ref{Eq:QuadraticProduct}).  This reveals the four roots to be
\begin{align}
\ell_{1} = \frac{-\sqrt{\xi_{1}}+\sqrt{-\xi_{1}+2b/\sqrt{\xi_{1}}}}{2}, \\
\ell_{2} = \frac{-\sqrt{\xi_{1}}-\sqrt{-\xi_{1}+2b/\sqrt{\xi_{1}}}}{2}, \\
\ell_{3} = \frac{\sqrt{\xi_{1}}+\sqrt{-\xi_{1}-2b/\sqrt{\xi_{1}}}}{2}, \\
\ell_{4} = \frac{\sqrt{\xi_{1}}-\sqrt{-\xi_{1}-2b/\sqrt{\xi_{1}}}}{2}.
\end{align}
Recalling that $b$ and $\xi_{1}$ are positive real, $\ell_{3}$ and $\ell_{4}$ are clearly complex, and therefore not of interest to us.  Noting that 
\begin{equation}
\xi_{1}^{3} = A^{3} + B^{3} + 3AB(A+B)  = b^{2} - 4c\xi_{1}, 
\label{Eq:xiCubed}
\end{equation} 
we can see that $A^{3}+B^{3}\equiv b^{2} >\xi_{1}^{3}$, hence
$4b^{2} > \xi_{1}^{3}$
and thus
$2b/\sqrt{\xi_{1}} > \xi_{1}$. Roots $\ell_{1}$ and $\ell_{2}$ are therefore real, but $\ell_{2}$ is guaranteed negative. However, from Eq.\ (\ref{Eq:xiCubed}) it also follows that 
\begin{equation}
\begin{split}
b>\xi_{1} \sqrt{\xi_{1}} \Rightarrow & 2b/\sqrt{\xi_{1}} > 2\xi_{1}
\Rightarrow  2b/\sqrt{\xi_{1}} -\xi_{1} > \xi_{1}
\\
\Rightarrow & \sqrt{-\xi_{1}+ 2b/\sqrt{\xi_{1}}} > \sqrt{\xi_{1}}.
\end{split}
\end{equation}
Hence $\ell_{1}$ is guaranteed positive real, and is the only solution of interest.

Thus, the single positive real root of Eq.\ (\ref{Eq:BaseQuartic}) is
\begin{equation}
\ell = \frac{\chi^{1/2}b^{1/3}}{2^{7/6}}
\left\{
\left[
\left(
\frac{2}{\chi}
\right)^{3/2}
-1
\right]^{1/2}
-1
\right\},
\label{Eq:EllValue}
\end{equation}
with
\begin{equation}
\chi =
\left\{
1 + 
\left[
1 + 
\frac{(c/3)^{3}}{(b/4)^{4}}
\right]^{1/2}
\right\}^{1/3}
+
\left\{
1 - 
\left[
1 + 
\frac{(c/3)^{3}}{(b/4)^{4}}
\right]^{1/2}
\right\}^{1/3},
\label{Eq:ChiValue}
\end{equation}
and where values of all fractional powers are taken to be real, and positive when a positive root exists.

\end{appendix}

 
%

\end{document}